\documentclass[sigconf,screen,authorversion]{acmart}

\AtBeginDocument{%
  \providecommand\BibTeX{{%
    \normalfont B\kern-0.5em{\scshape i\kern-0.25em b}\kern-0.8em\TeX}}}

\begin{document}

\title[Code Sophistication]{Code Sophistication}
\subtitle{From Code Recommendation to Logic Recommendation}

\author{Jessie Galasso}
\email{jessie.galasso-carbonnel@umontreal.ca}
\affiliation{%
  \institution{Université de Montréal, DIRO}
  \country{Canada}
}

\author{Michalis Famelis}
\email{michalis.famelis@umontreal.ca}
\affiliation{%
  \institution{Université de Montréal, DIRO}
  \country{Canada}
}

\author{Houari Sahraoui}
\email{houari.sahraoui@umontreal.ca}
\affiliation{%
  \institution{Université de Montréal, DIRO}
  \country{Canada}
}

\renewcommand{\shortauthors}{Galasso, Famelis and Sahraoui}
\begin{abstract}

A typical approach to programming is to first code the main execution scenario, and then focus on filling out alternative behaviors and corner cases.
But, almost always, there exist unusual conditions that trigger atypical behaviors, which are hard to predict in program specifications, and are thus often not coded.
In this paper, we consider the problem of detecting and recommending such missing  behaviors, a task that we call \textit{code sophistication}.
Previous research on coding assistants usually focuses on recommending code fragments based on specifications of the intended behavior.
In contrast, code sophistication happens in the absence of a specification, aiming to
help developers complete the logic of their programs with
missing and unspecified behaviors.
We outline the research challenges to this problem and
present early results showing how program logic can be completed  by
leveraging code structure and information about the usage of input parameters.

\end{abstract}


\settopmatter{printacmref=false}
\setcopyright{none}

\maketitle

\section{Introduction} 

When developing a program to solve a task, developers usually begin by coding typical scenarios, i.e., by implementing the behaviors that should be followed under normal conditions.
Subsequently,
they concentrate on atypical scenarios (e.g., failures, corner cases) by expanding the program with the appropriate alternative behaviors.  
Because these atypical scenarios correspond to unusual conditions, they may be omitted from specification documents~\cite{raghavan2004dex}, and are difficult to predict  during implementation~\cite{sidiroglou2015automatic}.
Even taking into account testing, such behaviors are difficult to detect as they are rarely triggered during executions~\cite{hemmati2015effective,myers2011art}.  
Unsurprisingly, they are one of the largest source of software defects~\cite{chang2007finding,duraes2006emulation,li2006have,raghavan2004dex}. 
It is thus crucial to help developers uncover such omitted scenarios and handle them by coding the appropriate corresponding program behaviors. 
These alternative behaviors may include error handling, variable assignations, early returns or more complex operations.

In this paper, we outline the problem of 
recommending unspecified behaviors to handle omitted scenarios, which we call \textit{code sophistication},
and present the results of our initial attempt to address it. 
The goal of code sophistication is to recommend missing parts of a program; we argue that finding missing behaviors cannot be properly handled by existing code recommendation approaches (see Section~\ref{sec:relatedwork}).
Such approaches typically rely on the existence of some kind of specification of what the program should be doing, and use it to suggest appropriate code fragments. 
But in the case of code sophistication, no specification of the missing behaviors can be assumed.
Thus the scope of code sophistication goes beyond offering developers assistance about \emph{how} to code but also assistance about \emph{what} to code to complete the logic of their program.
We propose an approach to leverage the knowledge embedded in large repositories of existing code to learn how to complete the logic of programs. 
We then present a preliminary implementation with promising results that allows
using this knowledge to infer potential missing behaviors in new programs.

The paper is organized as follows:
We cover related code recommendation techniques in Section~\ref{sec:relatedwork}.
We define and motivate the problem of code sophistication, and outline the research directions for addressing it in Section~\ref{sec:soph}.
We present our preliminary approach and results in 
Section~\ref{sec:evaluation},
and conclude in Section~\ref{sec:futurework}.

\section{Related Work}
\label{sec:relatedwork}

We can put techniques for assisting developers into completing their programs into four categories: code completion, program synthesis, automatic program repair, and code search. We explain why these techniques do not address code sophistication.

\textbf{Code completion} focuses on predicting the end of a token or the next tokens at a particular location in the code by relying on the code already written~\cite{allamanis2018survey}.
Source code language models extracted from large code bases represent probability distributions over sequences of tokens;
they were used to predict the next tokens given the previous ones~\cite{hindle2016naturalness,tu2014localness,hellendoorn2017deep}, identifiers~\cite{bhoopchand2016learning,li2017code},  function-calls~\cite{weyssow2020combining} or to infer meaningful code units (e.g., method call, binary expression)~\cite{nguyen2013statistical}.
Similar methods were used for more advanced predictions, such as API calls~\cite{bruch2009learning,proksch2015intelligent}, sequence of API invocations~\cite{raychev2014code} or even syntactic templates for API usages~\cite{nguyen2015graph}.
Code completion approaches do not rely on some explicit specification of what parts of the code should be recommended.
However they usually leverage low level syntactic and lexical features that are not sufficient to infer the logic of a program, much less to complete it, which is the aim of code sophistication.
The approach closest to sophistication is the work on inferring syntactic templates~\cite{nguyen2015graph} but it is restricted to specific API usages.

\textbf{Program synthesis} is about "\textit{discovering programs that realize user intent expressed in the form of some specifications}"~\cite{gulwani2010dimensions}.
The latter may be logical specifications, descriptions in natural language, input-output examples or test cases.
Early approaches relied on extensive specifications ~\cite{green1981application}, but more recent inductive approaches rather aim at producing generalization from partial specifications~\cite{kitzelmann2009inductive}.
Inductive program synthesis is usually seen as a search problem, where the goal is to explore the set of programs to find the one corresponding to the provided specification.
Approaches based on deep learning extract knowledge from large code bases, for instance to
 generate domain specific programs based on input/output examples~\cite{scott2015neural,gulwani2012spreadsheet}.
Other approaches rely on neural machine translation to translate functionality descriptions in natural language  into executable code in domain specific languages~\cite{zhong2017seq2sql}, general purpose languages for API usage sequences~\cite{gu2016deep}, or more generalized purpose~\cite{DBLP:conf/acl/LingBGHKWS16,DBLP:conf/acl/YinN17}.
\textbf{Automatic program repair} is a sub-field of program synthesis defined as "\textit{the transformation of an unacceptable behavior of a program into a acceptable one according to a specification}"~\cite{monperrus2018automatic}.
The goal of automatic program repair is to recommend appropriate patches to modify the part of the program responsible of an identified defect.
Some approaches use repair templates to address specific classes of defects~\cite{kim2013automatic,long2015staged,bader2019getafix}, while others rely on genetic programming and mutations to find appropriate patches~\cite{weimer2009automatically,arcuri2011evolutionary}.
Code Phage~\cite{sidiroglou2015automatic} is a system transferring correct code from programs passing a set of test cases to programs in which defects were detected.
Program synthesis and program repair approaches rely on specifications that characterize the missing behavior, such as test cases and textual descriptions.
Code recommendation is then akin to suggesting code fragments implementing an outlined behavior. In other words, it helps developers determine \emph{how} to code for a particular scenario.
On the contrary, code sophistication aims to suggest missing conditional paths corresponding to unspecified behaviors, relying only on the program under development. 
Note that some approaches in automatic program repair generate patches without relying on specifications~\cite{hata2018learning,tufano2019empirical,bader2019getafix}; however, they only target specific classes of defects.

When facing unfamiliar programming tasks, developers often seek code snippet examples; 
\textbf{code search} aims to suggest such relevant snippets. 
Most code search approaches are based on information retrieval techniques to find relevant code snippets depending on a query either formulated directly by the developer or inferred from the code~\cite{ai2019sensory}.
Some approaches focus on recommending specific types of snippets, such as framework usages~\cite{holmes2005using}, Java methods~\cite{ai2019sensory}, exception handling examples~\cite{rahman2014use}, auxiliary functionalities~\cite{lemos2011test} or API usages~\cite{raghothaman2016swim}, while others are general purpose~\cite{takuya2011spontaneous,jiang2016rosf,reiss2009semantics,jiang2018semantics,luan2019aroma,kim2018facoy}.
Approaches inferring a query from the code can be based on tokens and/or statements similarity~\cite{ai2019sensory,takuya2011spontaneous},  structural code features~\cite{holmes2005using,rahman2014use,kim2018facoy,luan2019aroma}, or lexical ones~\cite{rahman2014use}.
Code search approaches which are the closest to code sophistication are the ones in which no query is formulated directly and which rely only on the code under development to suggest code fragments. These approaches search for similar fragments and use the provided program as a specification of the intended behavior.
In contrast, code sophistication focuses on completing the logic of a program with behaviors that were missed. 
These behaviors are by definition assumed to be absent from the code. 
Code search approaches are thus not applicable for sophistication.

\section{Code Sophistication}
\label{sec:soph}

In this section, we first lay down and motivate the problem of code
sophistication. 
We then give directions for addressing it.

\subsection{Definitions and Motivations}

The set of behaviors of a program, sometimes referred as the \textit{program's logic}, can be outlined through execution paths, pseudocode, or use cases.
Execution paths offer an interesting perspective in our case, 
because we can think of each path as one behavior, depending on the conditions that delimit in which scenario it is executed~\cite{king1976symbolic}.
In other words, the choice of an execution  path is determined by the program’s inputs and the conditions they satisfy.  
Those inputs may be explicit (e.g., parameters, attributes) or implicit (e.g., time). 
Each conditional path thus defines how to behave in a scenario circumscribed by the values of the inputs.
Common scenarios correspond to combinations of input values that are typically observed, whereas atypical scenarios correspond to uncommon combinations which happen rarely.
\emph{Recommending alternative behaviors to handle omitted scenarios can thus be defined as suggesting missing conditional paths targeting specific combinations of input's values.} 

In software testing, missing conditional paths are known as a particular class of defects which do not reside in the written code, but in the absence of a particular fragment~\cite{myers2011art}. 
Several studies have analyzed bug fixes of large  projects and report that missing paths are a particularly abundant class of defects~\cite{raghavan2004dex,duraes2006emulation,li2006have,chang2007finding},  known to be hard to predict and detect.
In \textit{The Art of Software Testing}~\cite{myers2011art}, Myers et al. report that exhaustive path testing, a common approach to test the logic of a program, might not uncover errors caused by missing paths.
Hemmati showed that indeed the largest category of errors that remain undetected by code coverage criteria are "faults of omission" related to missing conditional paths~\cite{hemmati2015effective}.
Chen et al. studied dormant bugs, i.e., bugs introduced in a version of a system and not reported before after the release of the next version~\cite{chen2014empirical}.
They found that 52\% of dormant bugs are related to corner cases and control flow, while only 11\% of the non-dormant bugs concern these categories, suggesting they take longer to be exposed.
The reasons for this are that scenarios which are not present in the requirements are difficult to detect due to a lack of "local clues about the omission"~\cite{chang2008discovering} or because they necessitate particular conditions to be triggered~\cite{raghavan2004dex}. These suggest that missing conditional paths indeed correspond to omitted scenarios.

\begin{figure*}[bht]
    \centering
    \includegraphics[width=.8\linewidth]{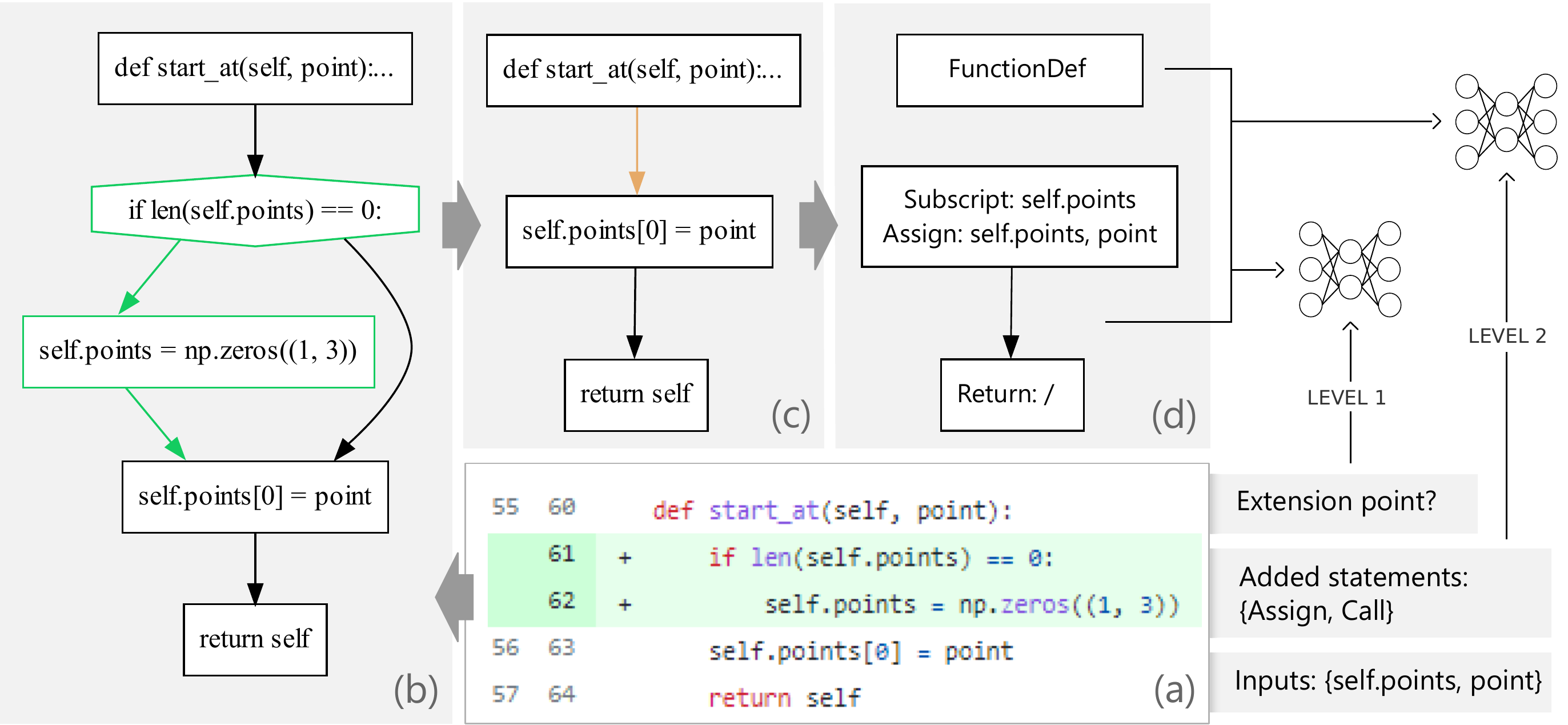}
    \caption{Processing a commit's method\protect\footnotemark~to detect extension points (level 1) and characterize missing behaviors (level 2).
    }
    \label{fig:experiments}
\end{figure*}

To sum up, the literature provides evidence that missing conditional paths corresponding to omitted scenarios are one of the most occurring class of defects found in software projects, and are particularly difficult to detect by both manual and automated software testing approaches.
Detecting and patching missing behaviors is thus an important and challenging issue.
In addition, we showed in Section~\ref{sec:relatedwork} that existing code recommendation approaches, while providing valuable avenues to investigate, are not sufficient to recommend missing behaviors in the general case.
In what follows, we provide directions for addressing this problem.

\subsection{Toward Logic Recommendation}

We hypothesize that knowledge about  programs' logic can be derived from available code in large project repositories.
More specifically, we aim at learning \textit{sophistication patterns} from recurring code changes across  project histories.
We consider that commits adding conditional paths, as illustrated in the Python method in Fig.~\ref{fig:experiments} (a), are good candidates to learn sophistication patterns.
Such patterns should characterize \textit{alternative behaviors} (i.e., what should be added) and their \textit{context} (i.e., where and when it should be added).

To achieve generalizability of sophistication patterns across different projects and domains, it is necessary to work with high level characterization of behaviors, independent from implementation concerns. 
For example, consider two applications, one for the reserving travel tickets, one for banking. The reservation application tests if the date of departure is before the date of return; the banking application tests if the amount of money to be withdrawn exceeds the bank balance. 
Despite having domain specific implementations, these behaviors are logically similar: they both raise an error if one input value is superior to another. 
Their characterization must be at a sufficient level of abstraction to reflect this similarity.

\footnotetext{Original commit: \url{https://github.com/3b1b/manim/commit/a9f620e2508cf9de4f5a405449673d4d50804d80}}

Because a program’s logic pertains to its structure, leveraging \textit{structural information} is crucial to depict the context of alternative behaviors.
We are currently investigating the use of structural representations of the code, notably the control flow graph (CFG).
A CFG is a graph representing the flow (directed edges) between the statements (nodes) of a program, portraying its different execution paths, as shown in Fig.~\ref{fig:experiments} (b) for the same Python method.
The nodes and edges in green represent the path added in the commit.
Predicate nodes have two outgoing edges and divide their ingoing path into two paths which depend on a condition.
Adding a new path in a program entails adding a predicate node in its corresponding CFG: thus, we consider that each edge of a CFG is a potential \textit{extension point}, on which a predicate node can be added to divide the current path and add an alternative behavior.
In Fig~\ref{fig:experiments} (b), we can see that the added predicate node (represented by an hexagon) splits the edge coming from the first node to insert a new path: this edge is an extension point.
Also, because the executed behaviors depend on the inputs of the program, we hypothesize that conditional paths may depend on the \textit{usage of inputs} through the program, and we take these usages into account when describing the context in sophistication patterns.

Based on these assumptions, we outline the problem of recommending missing behaviors as three levels of increasing complexity. 
\textit{(Level 1:)} Suggest possible extension points in a program.
Extension points show where the current control flow should be diverted when a certain condition is met to introduce a new behavior. 
\textit{(Level 2:)} Given an extension point in the program, suggest a high level description characterizing the potentially missing behavior.
Instead of suggesting code fragments representing specific implementation of a behavior, recommended solutions must rather characterize this behavior with a sufficient level of abstraction to be independent of domain specific implementation concerns. 
\textit{(Level 3:)} Given an extension point and a characterization of the missing behavior, suggest code templates of proposed implementations of the missing behavior.

\section{Implementation and Validation}
\label{sec:evaluation}

We propose a preliminary implementation of a code sophistication, based on Graph Convolutional Networks. We present early results for Levels 1 and 2: detecting potential extension points and characterizing missing behaviors.
Figure~\ref{fig:experiments} presents an overview of the proposed approach.

\subsection{Collecting and Encoding Data}

The change history of a software project is an important indicator of the defects which were uncovered and fixed by developers.  
To gather information about patterns of program logic completion, we focused on commits adding conditional statements, as they can show where a developer added a conditional path that was missing~\cite{raghavan2004dex}.
We analyzed all commits in the history of 250 Github projects written in Python, for a total of 1.2M commits, and extract methods in which an \texttt{if} statement block was added, as illustrated in  Fig.~\ref{fig:experiments} (a).
To leverage structural information in the extracted methods, we construct their control flow graph (CFG), as shown in  Fig.\ref{fig:experiments} (b). 
In each CFG, we  removed the nodes corresponding to the added path (Fig~\ref{fig:experiments} (c)): the pruned CFG thus corresponds to the CFG of the method before the commit.
We keep track of the edge corresponding to the extension point (represented in orange in Fig~\ref{fig:experiments} (c)), as well as the statements present in the path added by the commit (here an \texttt{Assign} and a \texttt{Call}).
Finally, we annotate each node of the CFG with the usage of the inputs, as illustrated in Fig~\ref{fig:experiments} (d).
We considered two types of inputs: the method's parameters and the attributes used in its body.
In our example, the method has two inputs: the parameter \texttt{point} and the attribute \texttt{self.points}.
We characterize the usage of inputs by the type of statements in which they appear.
In the middle node of Fig.~\ref{fig:experiments} (c), \texttt{self.points} is used in \texttt{Assign} and \texttt{Subscript} statements, and \texttt{point} in an \texttt{Assign} statement.
The \texttt{Return} statement of the bottom node does not use any input.
For each method, we thus have a CFG annotated with the usage of inputs, where one of its edges is identified as an extension point, associated with the statements from the removed path.

\subsection{Implementation of Learning Models}

For the two experiments we built classification models taking, as input, a CFG annotated with input usages.
We used the  StellarGraph library~\cite{StellarGraph} to build the graph classification models based on the graph convolutional layers from Kipf and Welling~\cite{kipf2016semi}.
Because we aim to represent an edge of the graph, we divide the CFG in two parts (before and after the tagged edge, representing the structured inputs' usage before and after the extension point, as shown in Fig~\ref{fig:experiments} (d)) and feed the models with these two sub-graphs.

To \textbf{detect extension points} (level 1), we trained a binary classifier to detect if an edge of a given CFG is an extension point.
For positive examples, we used the extracted graphs for which we know the edge on which a path was added.
For negative examples, we mined methods for which commits added lines that do not correspond to new paths. 
We created the same annotated CFGs, but with edges which did not correspond to extension points.
We evaluate the model with  the following metrics: accuracy, precision, recall, F1 measure and area under the curve (AUC). 
The AUC score measures the probability that our classifier will rank an edge corresponding to an extension point higher that an edge which does not.

To \textbf{characterize the missing behavior} (level 2), we trained a multi-label classifier to identify the types of statements that should be used in the behavior to be added at a given extension point. 
We selected 8 recurring types of statements as defined in the Python AST module~\footnote{\url{https://docs.python.org/3/library/ast}, accessed in July 2021} as our classes -- \texttt{Return}, \texttt{Assign}, \texttt{AugAssign}, \texttt{Raise}, \texttt{If}, \texttt{Call}, \texttt{Subscript} and \texttt{BinOp} -- and labeled each method with the statements appearing in the commit added lines.
We compared the results with two state-of-the-art models for code completion in Python, GraphCodeBERT~\cite{guo2020graphcodebert} and CodeGPT~\cite{lu2021codexglue}.
Because these two models do not suggest types of statements, we generate for each method the same number of tokens added in the commit. 
We then extracted the types of statements from the models' suggestions to obtain a baseline for comparison.
We evaluate the three models with the following metrics: AUC, macro and micro F1, and Hamming loss. 
AUC scores, in the multilabel case, represent an average of the AUC of each class.
Macro F1 is the average of F1 measure for each class, while micro F1 is the F1 measure computed on all examples and for all classes.
The Hamming loss represents how many times a pair edge-label is misclassified: the lower the loss the better.

\begin{table}[t]
    \centering
    \begin{tabular}{rrrrr}
    \hline
         Accuracy & AUC & Precision & Recall & F1 \\
    \hline
    .730 & .817 & .727 & .717 & .721  \\
    \hline
    \end{tabular}
    \caption{Results for the detection of extension points.}
    \label{tab:res_level1}
\vspace{-.6cm}
\end{table}

\begin{table}[t]

    \centering
    \begin{tabular}{lrrr}
    \hline
         Metrics & CC(CodeGPT) & CC(GraphCodeBERT) & \textbf{CS}  \\
    \hline
    AUC & .564 & .561 & \textbf{.750}  \\
    Macro F1 & .309 & .072 & \textbf{.358}\\
    Micro F1 & .333 & .095 & \textbf{.397}\\
    Hamming loss & .308  & .264 & \textbf{.237}\\
    \hline
    \end{tabular}
    \caption{Results for the characterization of  behaviors.}
    \label{tab:res_level2}
\vspace{-.4cm}
\end{table}

\subsection{Results}
For the two experiments, we performed a 5-fold cross validation: the presented results are the averages for the 5 folds.
Table~\ref{tab:res_level1} presents the results for the  model detecting extension points.
A classifier is often considered suitable if its AUC score is above 0.7~\cite{romano2011using}. 
Table~\ref{tab:res_level2} compares the results obtained with the two state-of-the-art models for code completion (denoted CC) and our multilabel classifier for code sophistication (denoted CS). We can see that our model provides better results than the two baselines.

In both cases, we have encouraging results suggesting that a) knowledge about programs' logic can be learned from code repositories and b) structural code information and input's usage are adequate to identify extension points and characterize missing behaviors, without relying on specifications.
\section{Conclusion and Future Work}
\label{sec:futurework}
We defined and motivate the problem of code sophistication, i.e., completing programs with missing behaviors that were neither specified nor predicted. 
We discussed how existing code recommendation approaches are not suitable for this problem and proposed an approach to recommending appropriate behaviors that relies on knowledge learned from large code repositories. 
We presented early results demonstrating we could detect extension points and characterize missing behaviors by learning from the structure and the usage of inputs from commits adding conditional paths.

Our next steps are to study in detail the added lines in the commits of our dataset to provide a characterization of the missing alternative behaviors, with descriptions independent from implementation and domain concerns.
This knowledge is crucial to better understand what parts of program logic we should aim to infer for code sophistication.
We are currently investigating the approaches used for related tasks, especially code completion and code search, with the goal of adapting and extending the more relevant ones to recommend alternative behaviors and code templates (\textit{Level 3}).
This will help us further determine which code properties are decisive to detect and infer missing behaviors.
In the long term, we aim to study how code sophistication could be efficiently carried out by recommendation systems to successfully assist developers into handling omitted scenarios.

\bibliographystyle{ACM-Reference-Format}
\bibliography{main.bib}


\begin{thebibliography}{55}


\ifx \showCODEN    \undefined \def \showCODEN     #1{\unskip}     \fi
\ifx \showDOI      \undefined \def \showDOI       #1{#1}\fi
\ifx \showISBNx    \undefined \def \showISBNx     #1{\unskip}     \fi
\ifx \showISBNxiii \undefined \def \showISBNxiii  #1{\unskip}     \fi
\ifx \showISSN     \undefined \def \showISSN      #1{\unskip}     \fi
\ifx \showLCCN     \undefined \def \showLCCN      #1{\unskip}     \fi
\ifx \shownote     \undefined \def \shownote      #1{#1}          \fi
\ifx \showarticletitle \undefined \def \showarticletitle #1{#1}   \fi
\ifx \showURL      \undefined \def \showURL       {\relax}        \fi
\providecommand\bibfield[2]{#2}
\providecommand\bibinfo[2]{#2}
\providecommand\natexlab[1]{#1}
\providecommand\showeprint[2][]{arXiv:#2}

\bibitem[\protect\citeauthoryear{Ai, Huang, Li, Zhou, and Yu}{Ai
  et~al\mbox{.}}{2019}]%
        {ai2019sensory}
\bibfield{author}{\bibinfo{person}{Lei Ai}, \bibinfo{person}{Zhiqiu Huang},
  \bibinfo{person}{Weiwei Li}, \bibinfo{person}{Yu Zhou}, {and}
  \bibinfo{person}{Yaoshen Yu}.} \bibinfo{year}{2019}\natexlab{}.
\newblock \showarticletitle{Sensory: Leveraging code statement sequence
  information for code snippets recommendation}. In
  \bibinfo{booktitle}{\emph{43rd Ann. Computer Software and Applications
  Conference}}, Vol.~\bibinfo{volume}{1}. IEEE, \bibinfo{pages}{27--36}.
\newblock


\bibitem[\protect\citeauthoryear{Allamanis, Barr, Devanbu, and
  Sutton}{Allamanis et~al\mbox{.}}{2018}]%
        {allamanis2018survey}
\bibfield{author}{\bibinfo{person}{Miltiadis Allamanis},
  \bibinfo{person}{Earl~T Barr}, \bibinfo{person}{Premkumar Devanbu}, {and}
  \bibinfo{person}{Charles Sutton}.} \bibinfo{year}{2018}\natexlab{}.
\newblock \showarticletitle{A survey of machine learning for big code and
  naturalness}.
\newblock \bibinfo{journal}{\emph{ACM Computing Surveys (CSUR)}}
  \bibinfo{volume}{51}, \bibinfo{number}{4} (\bibinfo{year}{2018}),
  \bibinfo{pages}{1--37}.
\newblock


\bibitem[\protect\citeauthoryear{Arcuri}{Arcuri}{2011}]%
        {arcuri2011evolutionary}
\bibfield{author}{\bibinfo{person}{Andrea Arcuri}.}
  \bibinfo{year}{2011}\natexlab{}.
\newblock \showarticletitle{Evolutionary repair of faulty software}.
\newblock \bibinfo{journal}{\emph{Applied soft computing}}
  \bibinfo{volume}{11}, \bibinfo{number}{4} (\bibinfo{year}{2011}),
  \bibinfo{pages}{3494--3514}.
\newblock


\bibitem[\protect\citeauthoryear{Bader, Scott, Pradel, and Chandra}{Bader
  et~al\mbox{.}}{2019}]%
        {bader2019getafix}
\bibfield{author}{\bibinfo{person}{Johannes Bader}, \bibinfo{person}{Andrew
  Scott}, \bibinfo{person}{Michael Pradel}, {and} \bibinfo{person}{Satish
  Chandra}.} \bibinfo{year}{2019}\natexlab{}.
\newblock \showarticletitle{Getafix: Learning to fix bugs automatically}.
\newblock \bibinfo{journal}{\emph{Proceedings of the ACM on Programming
  Languages}} \bibinfo{volume}{3}, \bibinfo{number}{OOPSLA}
  (\bibinfo{year}{2019}), \bibinfo{pages}{1--27}.
\newblock


\bibitem[\protect\citeauthoryear{Bhoopchand, Rockt{\"a}schel, Barr, and
  Riedel}{Bhoopchand et~al\mbox{.}}{2016}]%
        {bhoopchand2016learning}
\bibfield{author}{\bibinfo{person}{Avishkar Bhoopchand}, \bibinfo{person}{Tim
  Rockt{\"a}schel}, \bibinfo{person}{Earl Barr}, {and}
  \bibinfo{person}{Sebastian Riedel}.} \bibinfo{year}{2016}\natexlab{}.
\newblock \showarticletitle{Learning python code suggestion with a sparse
  pointer network}.
\newblock \bibinfo{journal}{\emph{arXiv preprint arXiv:1611.08307}}
  (\bibinfo{year}{2016}).
\newblock


\bibitem[\protect\citeauthoryear{Bruch, Monperrus, and Mezini}{Bruch
  et~al\mbox{.}}{2009}]%
        {bruch2009learning}
\bibfield{author}{\bibinfo{person}{Marcel Bruch}, \bibinfo{person}{Martin
  Monperrus}, {and} \bibinfo{person}{Mira Mezini}.}
  \bibinfo{year}{2009}\natexlab{}.
\newblock \showarticletitle{Learning from examples to improve code completion
  systems}. In \bibinfo{booktitle}{\emph{Proceedings of the 7th joint meeting
  of the European software engineering conference and the ACM SIGSOFT symposium
  on the foundations of software engineering}}. \bibinfo{pages}{213--222}.
\newblock


\bibitem[\protect\citeauthoryear{Chang, Podgurski, and Yang}{Chang
  et~al\mbox{.}}{2007}]%
        {chang2007finding}
\bibfield{author}{\bibinfo{person}{Ray-Yaung Chang}, \bibinfo{person}{Andy
  Podgurski}, {and} \bibinfo{person}{Jiong Yang}.}
  \bibinfo{year}{2007}\natexlab{}.
\newblock \showarticletitle{Finding what's not there: a new approach to
  revealing neglected conditions in software}. In
  \bibinfo{booktitle}{\emph{Proc. of the 2007 international symposium on
  Software testing and analysis}}. \bibinfo{pages}{163--173}.
\newblock


\bibitem[\protect\citeauthoryear{Chang, Podgurski, and Yang}{Chang
  et~al\mbox{.}}{2008}]%
        {chang2008discovering}
\bibfield{author}{\bibinfo{person}{Ray-Yaung Chang}, \bibinfo{person}{Andy
  Podgurski}, {and} \bibinfo{person}{Jiong Yang}.}
  \bibinfo{year}{2008}\natexlab{}.
\newblock \showarticletitle{Discovering neglected conditions in software by
  mining dependence graphs}.
\newblock \bibinfo{journal}{\emph{IEEE Transactions on Software Engineering}}
  \bibinfo{volume}{34}, \bibinfo{number}{5} (\bibinfo{year}{2008}),
  \bibinfo{pages}{579--596}.
\newblock


\bibitem[\protect\citeauthoryear{Chen, Nagappan, Shihab, and Hassan}{Chen
  et~al\mbox{.}}{2014}]%
        {chen2014empirical}
\bibfield{author}{\bibinfo{person}{Tse-Hsun Chen}, \bibinfo{person}{Meiyappan
  Nagappan}, \bibinfo{person}{Emad Shihab}, {and} \bibinfo{person}{Ahmed~E
  Hassan}.} \bibinfo{year}{2014}\natexlab{}.
\newblock \showarticletitle{An empirical study of dormant bugs}. In
  \bibinfo{booktitle}{\emph{Proceedings of the 11th Working Conference on
  Mining Software Repositories}}. \bibinfo{pages}{82--91}.
\newblock


\bibitem[\protect\citeauthoryear{Data61}{Data61}{2018}]%
        {StellarGraph}
\bibfield{author}{\bibinfo{person}{CSIRO's Data61}.}
  \bibinfo{year}{2018}\natexlab{}.
\newblock \bibinfo{title}{StellarGraph Machine Learning Library}.
\newblock
  \bibinfo{howpublished}{\url{https://github.com/stellargraph/stellargraph}}.
\newblock


\bibitem[\protect\citeauthoryear{Duraes and Madeira}{Duraes and
  Madeira}{2006}]%
        {duraes2006emulation}
\bibfield{author}{\bibinfo{person}{Joao~A Duraes} {and}
  \bibinfo{person}{Henrique~S Madeira}.} \bibinfo{year}{2006}\natexlab{}.
\newblock \showarticletitle{Emulation of software faults: A field data study
  and a practical approach}.
\newblock \bibinfo{journal}{\emph{Ieee transactions on software engineering}}
  \bibinfo{volume}{32}, \bibinfo{number}{11} (\bibinfo{year}{2006}),
  \bibinfo{pages}{849--867}.
\newblock


\bibitem[\protect\citeauthoryear{Green}{Green}{1981}]%
        {green1981application}
\bibfield{author}{\bibinfo{person}{Cordell Green}.}
  \bibinfo{year}{1981}\natexlab{}.
\newblock \showarticletitle{Application of theorem proving to problem solving}.
\newblock In \bibinfo{booktitle}{\emph{Readings in Artificial Intelligence}}.
  \bibinfo{publisher}{Elsevier}, \bibinfo{pages}{202--222}.
\newblock


\bibitem[\protect\citeauthoryear{Gu, Zhang, Zhang, and Kim}{Gu
  et~al\mbox{.}}{2016}]%
        {gu2016deep}
\bibfield{author}{\bibinfo{person}{Xiaodong Gu}, \bibinfo{person}{Hongyu
  Zhang}, \bibinfo{person}{Dongmei Zhang}, {and} \bibinfo{person}{Sunghun
  Kim}.} \bibinfo{year}{2016}\natexlab{}.
\newblock \showarticletitle{Deep API learning}. In
  \bibinfo{booktitle}{\emph{Proceedings of the 2016 24th ACM SIGSOFT
  International Symposium on Foundations of Software Engineering}}.
  \bibinfo{pages}{631--642}.
\newblock


\bibitem[\protect\citeauthoryear{Gulwani}{Gulwani}{2010}]%
        {gulwani2010dimensions}
\bibfield{author}{\bibinfo{person}{Sumit Gulwani}.}
  \bibinfo{year}{2010}\natexlab{}.
\newblock \showarticletitle{Dimensions in program synthesis}. In
  \bibinfo{booktitle}{\emph{12th Int. ACM SIGPLAN Symposium on Principles and
  practice of declarative programming}}. \bibinfo{pages}{13--24}.
\newblock


\bibitem[\protect\citeauthoryear{Gulwani, Harris, and Singh}{Gulwani
  et~al\mbox{.}}{2012}]%
        {gulwani2012spreadsheet}
\bibfield{author}{\bibinfo{person}{Sumit Gulwani}, \bibinfo{person}{William~R
  Harris}, {and} \bibinfo{person}{Rishabh Singh}.}
  \bibinfo{year}{2012}\natexlab{}.
\newblock \showarticletitle{Spreadsheet data manipulation using examples}.
\newblock \bibinfo{journal}{\emph{Commun. ACM}} \bibinfo{volume}{55},
  \bibinfo{number}{8} (\bibinfo{year}{2012}), \bibinfo{pages}{97--105}.
\newblock


\bibitem[\protect\citeauthoryear{Guo, Ren, Lu, Feng, Tang, Liu, Zhou, Duan,
  Svyatkovskiy, Fu, et~al\mbox{.}}{Guo et~al\mbox{.}}{2020}]%
        {guo2020graphcodebert}
\bibfield{author}{\bibinfo{person}{Daya Guo}, \bibinfo{person}{Shuo Ren},
  \bibinfo{person}{Shuai Lu}, \bibinfo{person}{Zhangyin Feng},
  \bibinfo{person}{Duyu Tang}, \bibinfo{person}{Shujie Liu},
  \bibinfo{person}{Long Zhou}, \bibinfo{person}{Nan Duan},
  \bibinfo{person}{Alexey Svyatkovskiy}, \bibinfo{person}{Shengyu Fu},
  {et~al\mbox{.}}} \bibinfo{year}{2020}\natexlab{}.
\newblock \showarticletitle{Graphcodebert: Pre-training code representations
  with data flow}.
\newblock \bibinfo{journal}{\emph{arXiv:2009.08366}} (\bibinfo{year}{2020}).
\newblock


\bibitem[\protect\citeauthoryear{Hata, Shihab, and Neubig}{Hata
  et~al\mbox{.}}{2018}]%
        {hata2018learning}
\bibfield{author}{\bibinfo{person}{Hideaki Hata}, \bibinfo{person}{Emad
  Shihab}, {and} \bibinfo{person}{Graham Neubig}.}
  \bibinfo{year}{2018}\natexlab{}.
\newblock \showarticletitle{Learning to generate corrective patches using
  neural machine translation}.
\newblock \bibinfo{journal}{\emph{arXiv:1812.07170}} (\bibinfo{year}{2018}).
\newblock


\bibitem[\protect\citeauthoryear{Hellendoorn and Devanbu}{Hellendoorn and
  Devanbu}{2017}]%
        {hellendoorn2017deep}
\bibfield{author}{\bibinfo{person}{Vincent~J Hellendoorn} {and}
  \bibinfo{person}{Premkumar Devanbu}.} \bibinfo{year}{2017}\natexlab{}.
\newblock \showarticletitle{Are deep neural networks the best choice for
  modeling source code?}. In \bibinfo{booktitle}{\emph{Proceedings of the 2017
  11th Joint Meeting on Foundations of Software Engineering}}.
  \bibinfo{pages}{763--773}.
\newblock


\bibitem[\protect\citeauthoryear{Hemmati}{Hemmati}{2015}]%
        {hemmati2015effective}
\bibfield{author}{\bibinfo{person}{Hadi Hemmati}.}
  \bibinfo{year}{2015}\natexlab{}.
\newblock \showarticletitle{How effective are code coverage criteria?}. In
  \bibinfo{booktitle}{\emph{2015 International Conference on Software Quality,
  Reliability and Security}}. IEEE, \bibinfo{pages}{151--156}.
\newblock


\bibitem[\protect\citeauthoryear{Hindle, Barr, Gabel, Su, and Devanbu}{Hindle
  et~al\mbox{.}}{2016}]%
        {hindle2016naturalness}
\bibfield{author}{\bibinfo{person}{Abram Hindle}, \bibinfo{person}{Earl~T
  Barr}, \bibinfo{person}{Mark Gabel}, \bibinfo{person}{Zhendong Su}, {and}
  \bibinfo{person}{Premkumar Devanbu}.} \bibinfo{year}{2016}\natexlab{}.
\newblock \showarticletitle{On the naturalness of software}.
\newblock \bibinfo{journal}{\emph{Commun. ACM}} \bibinfo{volume}{59},
  \bibinfo{number}{5} (\bibinfo{year}{2016}), \bibinfo{pages}{122--131}.
\newblock


\bibitem[\protect\citeauthoryear{Holmes and Murphy}{Holmes and Murphy}{2005}]%
        {holmes2005using}
\bibfield{author}{\bibinfo{person}{Reid Holmes} {and} \bibinfo{person}{Gail~C
  Murphy}.} \bibinfo{year}{2005}\natexlab{}.
\newblock \showarticletitle{Using structural context to recommend source code
  examples}. In \bibinfo{booktitle}{\emph{27th Int. Conf. on Software
  engineering}}. \bibinfo{pages}{117--125}.
\newblock


\bibitem[\protect\citeauthoryear{Jiang, Nie, Sun, Ren, Kong, Zhang, and
  Luo}{Jiang et~al\mbox{.}}{2016}]%
        {jiang2016rosf}
\bibfield{author}{\bibinfo{person}{He Jiang}, \bibinfo{person}{Liming Nie},
  \bibinfo{person}{Zeyi Sun}, \bibinfo{person}{Zhilei Ren},
  \bibinfo{person}{Weiqiang Kong}, \bibinfo{person}{Tao Zhang}, {and}
  \bibinfo{person}{Xiapu Luo}.} \bibinfo{year}{2016}\natexlab{}.
\newblock \showarticletitle{Rosf: Leveraging information retrieval and
  supervised learning for recommending code snippets}.
\newblock \bibinfo{journal}{\emph{Trans. on Services Computing}}
  \bibinfo{volume}{12} (\bibinfo{year}{2016}), \bibinfo{pages}{34--46}.
\newblock


\bibitem[\protect\citeauthoryear{Jiang, Chen, Zhang, Pei, Pan, and Zhang}{Jiang
  et~al\mbox{.}}{2018}]%
        {jiang2018semantics}
\bibfield{author}{\bibinfo{person}{Renhe Jiang}, \bibinfo{person}{Zhengzhao
  Chen}, \bibinfo{person}{Zejun Zhang}, \bibinfo{person}{Yu Pei},
  \bibinfo{person}{Minxue Pan}, {and} \bibinfo{person}{Tian Zhang}.}
  \bibinfo{year}{2018}\natexlab{}.
\newblock \showarticletitle{Semantics-based code search using input/output
  examples}. In \bibinfo{booktitle}{\emph{18th Int. Working Conf. on Source
  Code Analysis and Manipulation (SCAM)}}. IEEE, \bibinfo{pages}{92--102}.
\newblock


\bibitem[\protect\citeauthoryear{Kim, Nam, Song, and Kim}{Kim
  et~al\mbox{.}}{2013}]%
        {kim2013automatic}
\bibfield{author}{\bibinfo{person}{Dongsun Kim}, \bibinfo{person}{Jaechang
  Nam}, \bibinfo{person}{Jaewoo Song}, {and} \bibinfo{person}{Sunghun Kim}.}
  \bibinfo{year}{2013}\natexlab{}.
\newblock \showarticletitle{Automatic patch generation learned from
  human-written patches}. In \bibinfo{booktitle}{\emph{2013 35th International
  Conference on Software Engineering (ICSE)}}. IEEE, \bibinfo{pages}{802--811}.
\newblock


\bibitem[\protect\citeauthoryear{Kim, Kim, Bissyand{\'e}, Choi, Li, Klein, and
  Traon}{Kim et~al\mbox{.}}{2018}]%
        {kim2018facoy}
\bibfield{author}{\bibinfo{person}{Kisub Kim}, \bibinfo{person}{Dongsun Kim},
  \bibinfo{person}{Tegawend{\'e}~F Bissyand{\'e}}, \bibinfo{person}{Eunjong
  Choi}, \bibinfo{person}{Li Li}, \bibinfo{person}{Jacques Klein}, {and}
  \bibinfo{person}{Yves~Le Traon}.} \bibinfo{year}{2018}\natexlab{}.
\newblock \showarticletitle{FaCoY: a code-to-code search engine}. In
  \bibinfo{booktitle}{\emph{Proceedings of the 40th International Conference on
  Software Engineering}}. \bibinfo{pages}{946--957}.
\newblock


\bibitem[\protect\citeauthoryear{King}{King}{1976}]%
        {king1976symbolic}
\bibfield{author}{\bibinfo{person}{James~C King}.}
  \bibinfo{year}{1976}\natexlab{}.
\newblock \showarticletitle{Symbolic execution and program testing}.
\newblock \bibinfo{journal}{\emph{Commun. ACM}} \bibinfo{volume}{19},
  \bibinfo{number}{7} (\bibinfo{year}{1976}), \bibinfo{pages}{385--394}.
\newblock


\bibitem[\protect\citeauthoryear{Kipf and Welling}{Kipf and Welling}{2016}]%
        {kipf2016semi}
\bibfield{author}{\bibinfo{person}{Thomas~N Kipf} {and} \bibinfo{person}{Max
  Welling}.} \bibinfo{year}{2016}\natexlab{}.
\newblock \showarticletitle{Semi-supervised classification with graph
  convolutional networks}.
\newblock \bibinfo{journal}{\emph{arXiv preprint arXiv:1609.02907}}
  (\bibinfo{year}{2016}).
\newblock


\bibitem[\protect\citeauthoryear{Kitzelmann}{Kitzelmann}{2009}]%
        {kitzelmann2009inductive}
\bibfield{author}{\bibinfo{person}{Emanuel Kitzelmann}.}
  \bibinfo{year}{2009}\natexlab{}.
\newblock \showarticletitle{Inductive programming: A survey of program
  synthesis techniques}. In \bibinfo{booktitle}{\emph{International workshop on
  approaches and applications of inductive programming}}. Springer,
  \bibinfo{pages}{50--73}.
\newblock


\bibitem[\protect\citeauthoryear{Lemos, Bajracharya, Ossher, Masiero, and
  Lopes}{Lemos et~al\mbox{.}}{2011}]%
        {lemos2011test}
\bibfield{author}{\bibinfo{person}{Ot{\'a}vio Augusto~Lazzarini Lemos},
  \bibinfo{person}{Sushil Bajracharya}, \bibinfo{person}{Joel Ossher},
  \bibinfo{person}{Paulo~Cesar Masiero}, {and} \bibinfo{person}{Cristina
  Lopes}.} \bibinfo{year}{2011}\natexlab{}.
\newblock \showarticletitle{A test-driven approach to code search and its
  application to the reuse of auxiliary functionality}.
\newblock \bibinfo{journal}{\emph{Information and Software Technology}}
  \bibinfo{volume}{53}, \bibinfo{number}{4} (\bibinfo{year}{2011}),
  \bibinfo{pages}{294--306}.
\newblock


\bibitem[\protect\citeauthoryear{Li, Wang, Lyu, and King}{Li
  et~al\mbox{.}}{2017}]%
        {li2017code}
\bibfield{author}{\bibinfo{person}{Jian Li}, \bibinfo{person}{Yue Wang},
  \bibinfo{person}{Michael~R Lyu}, {and} \bibinfo{person}{Irwin King}.}
  \bibinfo{year}{2017}\natexlab{}.
\newblock \showarticletitle{Code completion with neural attention and pointer
  networks}.
\newblock \bibinfo{journal}{\emph{arXiv preprint arXiv:1711.09573}}
  (\bibinfo{year}{2017}).
\newblock


\bibitem[\protect\citeauthoryear{Li, Tan, Wang, Lu, Zhou, and Zhai}{Li
  et~al\mbox{.}}{2006}]%
        {li2006have}
\bibfield{author}{\bibinfo{person}{Zhenmin Li}, \bibinfo{person}{Lin Tan},
  \bibinfo{person}{Xuanhui Wang}, \bibinfo{person}{Shan Lu},
  \bibinfo{person}{Yuanyuan Zhou}, {and} \bibinfo{person}{Chengxiang Zhai}.}
  \bibinfo{year}{2006}\natexlab{}.
\newblock \showarticletitle{Have things changed now? An empirical study of bug
  characteristics in modern open source software}. In
  \bibinfo{booktitle}{\emph{Proceedings of the 1st workshop on Architectural
  and system support for improving software dependability}}.
  \bibinfo{pages}{25--33}.
\newblock


\bibitem[\protect\citeauthoryear{Ling, Blunsom, Grefenstette, Hermann,
  Kocisk{\'{y}}, Wang, and Senior}{Ling et~al\mbox{.}}{2016}]%
        {DBLP:conf/acl/LingBGHKWS16}
\bibfield{author}{\bibinfo{person}{Wang Ling}, \bibinfo{person}{Phil Blunsom},
  \bibinfo{person}{Edward Grefenstette}, \bibinfo{person}{Karl~Moritz Hermann},
  \bibinfo{person}{Tom{\'{a}}s Kocisk{\'{y}}}, \bibinfo{person}{Fumin Wang},
  {and} \bibinfo{person}{Andrew~W. Senior}.} \bibinfo{year}{2016}\natexlab{}.
\newblock \showarticletitle{Latent Predictor Networks for Code Generation}. In
  \bibinfo{booktitle}{\emph{Proceedings of the 54th Annual Meeting of the
  Association for Computational Linguistics, {ACL} 2016}}.
\newblock


\bibitem[\protect\citeauthoryear{Long and Rinard}{Long and Rinard}{2015}]%
        {long2015staged}
\bibfield{author}{\bibinfo{person}{Fan Long} {and} \bibinfo{person}{Martin
  Rinard}.} \bibinfo{year}{2015}\natexlab{}.
\newblock \showarticletitle{Staged program repair with condition synthesis}. In
  \bibinfo{booktitle}{\emph{10th Joint Meeting on Foundations of Software
  Engineering}}. \bibinfo{pages}{166--178}.
\newblock


\bibitem[\protect\citeauthoryear{Lu, Guo, Ren, Huang, Svyatkovskiy, Blanco,
  Clement, Drain, Jiang, Tang, et~al\mbox{.}}{Lu et~al\mbox{.}}{2021}]%
        {lu2021codexglue}
\bibfield{author}{\bibinfo{person}{Shuai Lu}, \bibinfo{person}{Daya Guo},
  \bibinfo{person}{Shuo Ren}, \bibinfo{person}{Junjie Huang},
  \bibinfo{person}{Alexey Svyatkovskiy}, \bibinfo{person}{Ambrosio Blanco},
  \bibinfo{person}{Colin Clement}, \bibinfo{person}{Dawn Drain},
  \bibinfo{person}{Daxin Jiang}, \bibinfo{person}{Duyu Tang}, {et~al\mbox{.}}}
  \bibinfo{year}{2021}\natexlab{}.
\newblock \showarticletitle{CodeXGLUE: A Machine Learning Benchmark Dataset for
  Code Understanding and Generation}.
\newblock \bibinfo{journal}{\emph{arXiv preprint arXiv:2102.04664}}
  (\bibinfo{year}{2021}).
\newblock


\bibitem[\protect\citeauthoryear{Luan, Yang, Barnaby, Sen, and Chandra}{Luan
  et~al\mbox{.}}{2019}]%
        {luan2019aroma}
\bibfield{author}{\bibinfo{person}{Sifei Luan}, \bibinfo{person}{Di Yang},
  \bibinfo{person}{Celeste Barnaby}, \bibinfo{person}{Koushik Sen}, {and}
  \bibinfo{person}{Satish Chandra}.} \bibinfo{year}{2019}\natexlab{}.
\newblock \showarticletitle{Aroma: Code recommendation via structural code
  search}.
\newblock \bibinfo{journal}{\emph{Proceedings of the ACM on Programming
  Languages}} \bibinfo{volume}{3}, \bibinfo{number}{OOPSLA}
  (\bibinfo{year}{2019}), \bibinfo{pages}{1--28}.
\newblock


\bibitem[\protect\citeauthoryear{Monperrus}{Monperrus}{2018}]%
        {monperrus2018automatic}
\bibfield{author}{\bibinfo{person}{Martin Monperrus}.}
  \bibinfo{year}{2018}\natexlab{}.
\newblock \showarticletitle{Automatic software repair: a bibliography}.
\newblock \bibinfo{journal}{\emph{ACM Computing Surveys (CSUR)}}
  \bibinfo{volume}{51}, \bibinfo{number}{1} (\bibinfo{year}{2018}),
  \bibinfo{pages}{1--24}.
\newblock


\bibitem[\protect\citeauthoryear{Myers, Sandler, and Badgett}{Myers
  et~al\mbox{.}}{2011}]%
        {myers2011art}
\bibfield{author}{\bibinfo{person}{Glenford~J Myers}, \bibinfo{person}{Corey
  Sandler}, {and} \bibinfo{person}{Tom Badgett}.}
  \bibinfo{year}{2011}\natexlab{}.
\newblock \bibinfo{booktitle}{\emph{The art of software testing}}.
\newblock \bibinfo{publisher}{John Wiley \& Sons}.
\newblock


\bibitem[\protect\citeauthoryear{Nguyen and Nguyen}{Nguyen and Nguyen}{2015}]%
        {nguyen2015graph}
\bibfield{author}{\bibinfo{person}{Anh~Tuan Nguyen} {and}
  \bibinfo{person}{Tien~N Nguyen}.} \bibinfo{year}{2015}\natexlab{}.
\newblock \showarticletitle{Graph-based statistical language model for code}.
  In \bibinfo{booktitle}{\emph{37th Int. Conf. on Software Engineering}},
  Vol.~\bibinfo{volume}{1}. IEEE, \bibinfo{pages}{858--868}.
\newblock


\bibitem[\protect\citeauthoryear{Nguyen, Nguyen, Nguyen, and Nguyen}{Nguyen
  et~al\mbox{.}}{2013}]%
        {nguyen2013statistical}
\bibfield{author}{\bibinfo{person}{Tung~Thanh Nguyen},
  \bibinfo{person}{Anh~Tuan Nguyen}, \bibinfo{person}{Hoan~Anh Nguyen}, {and}
  \bibinfo{person}{Tien~N Nguyen}.} \bibinfo{year}{2013}\natexlab{}.
\newblock \showarticletitle{A statistical semantic language model for source
  code}. In \bibinfo{booktitle}{\emph{Proceedings of the 9th Joint Meeting on
  Foundations of Software Engineering}}. \bibinfo{pages}{532--542}.
\newblock


\bibitem[\protect\citeauthoryear{Proksch, Lerch, and Mezini}{Proksch
  et~al\mbox{.}}{2015}]%
        {proksch2015intelligent}
\bibfield{author}{\bibinfo{person}{Sebastian Proksch},
  \bibinfo{person}{Johannes Lerch}, {and} \bibinfo{person}{Mira Mezini}.}
  \bibinfo{year}{2015}\natexlab{}.
\newblock \showarticletitle{Intelligent code completion with Bayesian
  networks}.
\newblock \bibinfo{journal}{\emph{ACM Transactions on Software Engineering and
  Methodology (TOSEM)}} \bibinfo{volume}{25}, \bibinfo{number}{1}
  (\bibinfo{year}{2015}), \bibinfo{pages}{1--31}.
\newblock


\bibitem[\protect\citeauthoryear{Raghavan, Rohana, Leon, Podgurski, and
  Augustine}{Raghavan et~al\mbox{.}}{2004}]%
        {raghavan2004dex}
\bibfield{author}{\bibinfo{person}{Shruti Raghavan}, \bibinfo{person}{Rosanne
  Rohana}, \bibinfo{person}{David Leon}, \bibinfo{person}{Andy Podgurski},
  {and} \bibinfo{person}{Vinay Augustine}.} \bibinfo{year}{2004}\natexlab{}.
\newblock \showarticletitle{Dex: A semantic-graph differencing tool for
  studying changes in large code bases}. In \bibinfo{booktitle}{\emph{20th Int.
  Conf. on Software Maintenance}}. IEEE, \bibinfo{pages}{188--197}.
\newblock


\bibitem[\protect\citeauthoryear{Raghothaman, Wei, and Hamadi}{Raghothaman
  et~al\mbox{.}}{2016}]%
        {raghothaman2016swim}
\bibfield{author}{\bibinfo{person}{Mukund Raghothaman}, \bibinfo{person}{Yi
  Wei}, {and} \bibinfo{person}{Youssef Hamadi}.}
  \bibinfo{year}{2016}\natexlab{}.
\newblock \showarticletitle{Swim: Synthesizing what i mean-code search and
  idiomatic snippet synthesis}. In \bibinfo{booktitle}{\emph{2016 IEEE/ACM 38th
  International Conference on Software Engineering (ICSE)}}. IEEE,
  \bibinfo{pages}{357--367}.
\newblock


\bibitem[\protect\citeauthoryear{Rahman and Roy}{Rahman and Roy}{2014}]%
        {rahman2014use}
\bibfield{author}{\bibinfo{person}{Mohammad~Masudur Rahman} {and}
  \bibinfo{person}{Chanchal~K Roy}.} \bibinfo{year}{2014}\natexlab{}.
\newblock \showarticletitle{On the use of context in recommending exception
  handling code examples}. In \bibinfo{booktitle}{\emph{2014 IEEE 14th
  International Working Conference on Source Code Analysis and Manipulation}}.
  IEEE, \bibinfo{pages}{285--294}.
\newblock


\bibitem[\protect\citeauthoryear{Raychev, Vechev, and Yahav}{Raychev
  et~al\mbox{.}}{2014}]%
        {raychev2014code}
\bibfield{author}{\bibinfo{person}{Veselin Raychev}, \bibinfo{person}{Martin
  Vechev}, {and} \bibinfo{person}{Eran Yahav}.}
  \bibinfo{year}{2014}\natexlab{}.
\newblock \showarticletitle{Code completion with statistical language models}.
  In \bibinfo{booktitle}{\emph{Proceedings of the 35th ACM SIGPLAN Conference
  on Programming Language Design and Implementation}}.
  \bibinfo{pages}{419--428}.
\newblock


\bibitem[\protect\citeauthoryear{Reiss}{Reiss}{2009}]%
        {reiss2009semantics}
\bibfield{author}{\bibinfo{person}{Steven~P Reiss}.}
  \bibinfo{year}{2009}\natexlab{}.
\newblock \showarticletitle{Semantics-based code search}. In
  \bibinfo{booktitle}{\emph{2009 IEEE 31st International Conference on Software
  Engineering}}. IEEE, \bibinfo{pages}{243--253}.
\newblock


\bibitem[\protect\citeauthoryear{Romano and Pinzger}{Romano and
  Pinzger}{2011}]%
        {romano2011using}
\bibfield{author}{\bibinfo{person}{Daniele Romano} {and}
  \bibinfo{person}{Martin Pinzger}.} \bibinfo{year}{2011}\natexlab{}.
\newblock \showarticletitle{Using source code metrics to predict change-prone
  java interfaces}. In \bibinfo{booktitle}{\emph{2011 27th IEEE international
  conference on software maintenance (ICSM)}}. IEEE, \bibinfo{pages}{303--312}.
\newblock


\bibitem[\protect\citeauthoryear{Scott and Freitas}{Scott and Freitas}{2015}]%
        {scott2015neural}
\bibfield{author}{\bibinfo{person}{Reed Scott} {and} \bibinfo{person}{N
  Freitas}.} \bibinfo{year}{2015}\natexlab{}.
\newblock \showarticletitle{Neural programmer-interpreters}. In
  \bibinfo{booktitle}{\emph{International Conference on Learning
  Representations}}.
\newblock


\bibitem[\protect\citeauthoryear{Sidiroglou-Douskos, Lahtinen, Long, and
  Rinard}{Sidiroglou-Douskos et~al\mbox{.}}{2015}]%
        {sidiroglou2015automatic}
\bibfield{author}{\bibinfo{person}{Stelios Sidiroglou-Douskos},
  \bibinfo{person}{Eric Lahtinen}, \bibinfo{person}{Fan Long}, {and}
  \bibinfo{person}{Martin Rinard}.} \bibinfo{year}{2015}\natexlab{}.
\newblock \showarticletitle{Automatic error elimination by horizontal code
  transfer across multiple applications}. In
  \bibinfo{booktitle}{\emph{Proceedings of the 36th ACM SIGPLAN Conference on
  Programming Language Design and Implementation}}. \bibinfo{pages}{43--54}.
\newblock


\bibitem[\protect\citeauthoryear{Takuya and Masuhara}{Takuya and
  Masuhara}{2011}]%
        {takuya2011spontaneous}
\bibfield{author}{\bibinfo{person}{Watanabe Takuya} {and}
  \bibinfo{person}{Hidehiko Masuhara}.} \bibinfo{year}{2011}\natexlab{}.
\newblock \showarticletitle{A spontaneous code recommendation tool based on
  associative search}. In \bibinfo{booktitle}{\emph{Proc. of the 3rd
  International Workshop on search-driven development: Users, infrastructure,
  tools, and evaluation}}. \bibinfo{pages}{17--20}.
\newblock


\bibitem[\protect\citeauthoryear{Tu, Su, and Devanbu}{Tu et~al\mbox{.}}{2014}]%
        {tu2014localness}
\bibfield{author}{\bibinfo{person}{Zhaopeng Tu}, \bibinfo{person}{Zhendong Su},
  {and} \bibinfo{person}{Premkumar Devanbu}.} \bibinfo{year}{2014}\natexlab{}.
\newblock \showarticletitle{On the localness of software}. In
  \bibinfo{booktitle}{\emph{22nd Int. Symp. on Foundations of Software
  Engineering}}. \bibinfo{pages}{269--280}.
\newblock


\bibitem[\protect\citeauthoryear{Tufano, Watson, Bavota, Penta, White, and
  Poshyvanyk}{Tufano et~al\mbox{.}}{2019}]%
        {tufano2019empirical}
\bibfield{author}{\bibinfo{person}{Michele Tufano}, \bibinfo{person}{Cody
  Watson}, \bibinfo{person}{Gabriele Bavota}, \bibinfo{person}{Massimiliano~Di
  Penta}, \bibinfo{person}{Martin White}, {and} \bibinfo{person}{Denys
  Poshyvanyk}.} \bibinfo{year}{2019}\natexlab{}.
\newblock \showarticletitle{An empirical study on learning bug-fixing patches
  in the wild via neural machine translation}.
\newblock \bibinfo{journal}{\emph{ACM Transactions on Software Engineering and
  Methodology (TOSEM)}} \bibinfo{volume}{28}, \bibinfo{number}{4}
  (\bibinfo{year}{2019}), \bibinfo{pages}{1--29}.
\newblock


\bibitem[\protect\citeauthoryear{Weimer, Nguyen, Le~Goues, and Forrest}{Weimer
  et~al\mbox{.}}{2009}]%
        {weimer2009automatically}
\bibfield{author}{\bibinfo{person}{Westley Weimer}, \bibinfo{person}{ThanhVu
  Nguyen}, \bibinfo{person}{Claire Le~Goues}, {and} \bibinfo{person}{Stephanie
  Forrest}.} \bibinfo{year}{2009}\natexlab{}.
\newblock \showarticletitle{Automatically finding patches using genetic
  programming}. In \bibinfo{booktitle}{\emph{2009 IEEE 31st International
  Conference on Software Engineering}}. IEEE, \bibinfo{pages}{364--374}.
\newblock


\bibitem[\protect\citeauthoryear{Weyssow, Sahraoui, Fr{\'e}nay, and
  Vanderose}{Weyssow et~al\mbox{.}}{2020}]%
        {weyssow2020combining}
\bibfield{author}{\bibinfo{person}{Martin Weyssow}, \bibinfo{person}{Houari
  Sahraoui}, \bibinfo{person}{Beno{\i}t Fr{\'e}nay}, {and}
  \bibinfo{person}{Beno{\i}t Vanderose}.} \bibinfo{year}{2020}\natexlab{}.
\newblock \showarticletitle{Combining Code Embedding with Static Analysis for
  Function-Call Completion}.
\newblock \bibinfo{journal}{\emph{arXiv preprint arXiv:2008.03731}}
  (\bibinfo{year}{2020}).
\newblock


\bibitem[\protect\citeauthoryear{Yin and Neubig}{Yin and Neubig}{2017}]%
        {DBLP:conf/acl/YinN17}
\bibfield{author}{\bibinfo{person}{Pengcheng Yin} {and} \bibinfo{person}{Graham
  Neubig}.} \bibinfo{year}{2017}\natexlab{}.
\newblock \showarticletitle{A Syntactic Neural Model for General-Purpose Code
  Generation}. In \bibinfo{booktitle}{\emph{55th Ann. Meet. of the Association
  for Computational Linguistics}}, \bibfield{editor}{\bibinfo{person}{Regina
  Barzilay} {and} \bibinfo{person}{Min{-}Yen Kan}} (Eds.).
  \bibinfo{publisher}{ACL}, \bibinfo{pages}{440--450}.
\newblock


\bibitem[\protect\citeauthoryear{Zhong, Xiong, and Socher}{Zhong
  et~al\mbox{.}}{2017}]%
        {zhong2017seq2sql}
\bibfield{author}{\bibinfo{person}{Victor Zhong}, \bibinfo{person}{Caiming
  Xiong}, {and} \bibinfo{person}{Richard Socher}.}
  \bibinfo{year}{2017}\natexlab{}.
\newblock \showarticletitle{Seq2sql: Generating structured queries from natural
  language using reinforcement learning}.
\newblock \bibinfo{journal}{\emph{arXiv preprint arXiv:1709.00103}}
  (\bibinfo{year}{2017}).
\newblock


\end{thebibliography}

\end{document}